\documentclass[twocolumn,showpacs,showkeys,preprintnumbers,amsmath,amssymb]{revtex4}

\usepackage[dvips]{graphicx}
\usepackage{dcolumn}
\usepackage{bm}

\begin{document}

\title{Turing structures in dc gas discharges}

\author{S. Popescu}

\affiliation{"Al I Cuza" University, 700506 Iasi, Romania}%

\begin{abstract}
Self-organized spatial plasma structures exhibit many similarities with
Turing structures obtained in biology and chemistry. Using an analytical
mesoscopic approach it is shown that plasma balls of fire belong to the same
class of Turing structures like the Brusselator. It is also mathematically
proved that the existence of these self-organized plasma structures is
related with a negative differential resistance. There are also established
the mathematical conditions which the negative differential resistance must
satisfy for obtaining a stationary ball of fire at the anode of a plasma
diode.
\end{abstract}

\maketitle

\section{Introduction}

In a paper published in 1952 \cite{Turing}, the British mathematician Alan
Mathison Turing suggested a possible connection between biologic structures
emerged during the morphogenesis process and chemical structures
spontaneously formed in reaction-diffusion (RD) systems. For 15 years that
paper did not have any echo among scientists, but starting with 1960s Ilya
Prigogine's School from Brussels, Belgium begun an intensive study of
Turing's idea \cite{Prigogine1}. The result was a theoretical chemical
model, known today by the name of Brusselator \cite{Prigogine2}. In spite of
these positive theoretical results, the first experimental evidences of
chemical Turing structures were obtained only in 1990-1991 by the groups of
Patrick de Kepper in Bordeaux, France \cite{Kepper} and of Qi Ouyang in
Austin, Texas, USA \cite{Ouyang}.

The theoretical approach of Turing structures involves the study of the RD
equations describing the system under study. The general method of treating
this kind of mathematical problem is outlined in \cite{Murray}, \cite{Koch}
and it implies the use of the tools of nonlinear dynamics. Another
mathematical and conceptual method for studying the Turing structures in
solid state, but also in plasma physics, is by considering them as
autosolitons \cite{Kerner}. Recently, with the help of numerical
simulations, there was obtained a large variety of spatial patterns and
structures in two and three dimensions, from spots and stripes, to lamellae
and spherical droplets \cite{Lepp}.

Almost 45 years after the publication of Turing's paper, a group of
physicists from M\"{u}nster University, Germany \cite{Astrov,Purwins}
have experimentally observed that, as a result of dc discharges in a
quasi-bidimensional geometry, self-organized luminous structures have been
formed, in the form of spots, striated filaments, hexagons and stripes. The
self-assembling of such patterns has been attributed to a Turing-type
scenario, analogous to that encountered in chemical systems \cite{Kerner,Astrov,Purwins}.

Developing some similar experiments with the above ones, in a different
experimental arrangement, but also in a bidimensional geometry, we analyzed
the succession of the physical processes at the basis of the current
filamentation in plasma systems \cite{Loz1}. The experiments proved that at
the origin of current filaments is the formation of a quasi-bidimensional
anode spot. Its emergence follows the same physical scenario as in the
three-dimensional case, where it is also known as "plasma ball of fire" \cite%
{Loz2}. In the attempt to search if plasma balls of fire are Turing-type
structures, our previous theoretical studies proved that taking into account
elementary quantum processes, like excitation and ionization, which are the
key ingredients for ball of fire formation, Turing-like structures cannot
arise in plasma \cite{Popescu}. In other words, the above study proved
once more that for describing the emergence of a self-organized structure
the mesoscopic approach is necessary, the microscopic one being inadequate.

The aim of this paper is to prove analytically, on the basis of a mesoscopic
approach (\textit{i.e.} starting from the equivalent electric circuit of the
plasma system), that the appearance of a ball of fire is associated with a
negative differential resistance. The present results also prove, for the
first time to the knowledge of the author, that the plasma ball of fire
belongs to the same class of Turing structures like the Brusselator:
activator - substrate depleted. There are also established the mathematical
conditions the negative differential resistance must satisfy for obtaining a
stationary ball of fire at the anode of a plasma diode.

\section{What are Turing structures?}

In his seminal paper \cite{Turing}, Turing showed that, in certain
conditions, a homogeneous biologic medium can undergo a spatial symmetry
breaking, giving birth to a spatial ordered structure. The present
understanding of the concept of Turing structure is mainly due to de
Kepper's group \cite{Kepper}: Turing structures are self-organized
stationary spatial structures appearing in dissipative systems and
correspond to the stationary stable solutions of the RD equations modelling
the studied physical system. Turing structures appear only in open systems 
\cite{Turing}, far from the thermodynamic equilibrium, where the essential
processes - reaction and diffusion - can be coupled. As already proved \cite%
{Murray}, one of the key processes for the appearance of spatial structures
is the autocatalytic one, \textit{i.e.} the self-enhancement of a chemical
species, called \textit{activator}. The other important species in the
appearance of Turing structures is called \textit{inhibitor}, or, in
different circumstances, \textit{substrate depleted}. Both, the activator
and the inhibitor (or substrate depleted), are referred to as \textit{%
intermediate species}. Finally, the rest of chemicals existent in the system
are called \textit{pool species} \cite{Koch}. The destabilization of the
stationary homogeneous state and the emergence of a Turing structure can be
realized only if the diffusivities of the intermediate species are different 
\cite{Vastano}. If the stationary homogeneous state of an RD system is
stable to small spatial perturbations in the absence of diffusion, but
unstable when the diffusion process is present \cite{Murray}, then an
instability will take place in the system. The result of this instability,
now called Turing instability, is the spatial structuralization of the
system. Compensating the difference between the reaction rates, the
diffusion makes the inhomogeneous state be stationary \cite{Prigogine1}.
Of course, this is valid as long as the system feeding is continuous,
otherwise the structures are transitory.

More recently, it was demonstrated that another necessary condition for the
appearance of Turing structures is the presence of cross-inhibition \cite%
{Szili}, that is an intermediate species inhibits the increase of another
intermediate species' concentration.

\section{Plasma balls of fire}

When a dc power supply is connected, through a load resistor, to a plasma
diode, the processes taking place in front of the anode are numerous, they
depending on the magnitude of the potential drop on the diode. The
experiments clearly show that, as the plasma system is gradually departed
from the thermodynamic equilibrium, the processes inside the diode become
strongly nonlinear.

Gradually increasing from zero the potential drop on the plasma diode, the
plasma system first behaves like an ohmic conductor, after that the static
I(V) characteristic becoming nonlinear \cite{Sandu1,Leu}. The
nonlinearity of the static I(V) characteristic is related with the quantum
processes of excitation and ionization taking place in front of the anode.
These processes give birth, in order, to an electron layer and a positive
ion layer, the later being localized between the electronic one and the
anode. The interactions between these layers are collective ones, the
electrostatic forces acting between large groups of electric opposite
charges. They lead to the self-assembling of a planar electric double layer
in front of the anode. Increasing further the potential drop on the plasma
diode, there will be a threshold potential drop for which the planar double
layer transits into a spherical one for minimizing the free energy of the
system. The spherical double layer, having the negative sheath at the
exterior, covers a nucleus consisting of a plasma enriched in positive ions
[Fig. 1(a)]. This luminous, beautifully colored, complex space charge
configuration, formed at the anode surface, is known by the name of anode
spot or plasma ball of fire \cite{Sandu1}. During the self-assembling of the
ball of fire, which is a spontaneous process, the static I(V) characteristic
displays a sudden jump, the current increasing abruptly, while the potential
drop on the diode is at its threshold value.

\begin{figure*}[tb]
\centering
\includegraphics{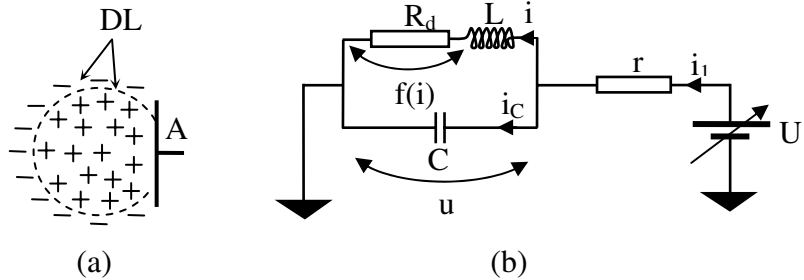}
\caption{\label{fig:Fig1} (a) Schematic representation of plasma ball of
fire. A - anode (disc electrode 3 cm in diameter), DL - electric double
layer, plasma parameters: Ar, pressure 10 mTorr, density 10$^{9}$ cm$^{-3}$
[18]; (b) Electric circuit of a plasma diode with a ball of fire on the
anode. \textit{U} is the variable voltage of the dc power supply, \textit{r}
is the load resistor, \textit{R}$_{d}$ is the dynamic resistance of the ball
of fire, \textit{L }its inductance, \textit{C} its capacitance, and \textit{u} the potential drop on the diode.}
\end{figure*}

Increasing the potential drop even slightly above the threshold value, the
ball of fire exists in a stationary state at the anode surface. Decreasing
now the potential drop on the diode, the ball of fire will exist in the
stationary state even for potential drops smaller than the threshold value.
This means nothing else than the existence of the ball of fire is displayed
in the static I(V) characteristic by a hysteresis loop, which is one of the
fingerprints of self-organization, proving that the structure has memory.
The stable branches of the hysteresis loop coexist, meaning that the
characteristic displays a region of bistability. Between these two stable
branches there is another one, unstable, on which the differential
resistance of the gaseous conductor is negative. Because the succession of
these three states has the form of the letter S, the negative differential
resistance is commonly called S-type negative differential resistance. This
negative differential resistance proves that the ball of fire acts as an
energy reservoir, allowing the structure to exist for a while, even for
worse external conditions than those needed for its emergence.

\section{Circuit theory for a plasma diode with a ball of fire at the anode
surface}

The experiments studying the emergence of the plasma ball of fire revealed
the following: the double layer bordering the self-organized structure has
an electric capacity \cite{Popescu2}; the ball of fire has an inductance
related with the inertial properties of the positive ions from its nucleus;
last but not least, the ball of fire is responsible for the hysteresis loop
appearance, as well as for electric oscillations in the anode circuit, which
means that it acts as a negative differential resistance \cite{Sandu1}.

Based on the above experimental results and choosing the experimental
arrangement such that only the anode part of the discharge to be important,
the simplest equivalent electric circuit for the plasma diode can be
established and it looks like that in Fig. 1(b). Because the ball of fire is
a nonlinear circuit element, the potential drop on its dynamic resistance is
given by a nonlinear function of the current, labeled by $f(i)$ in Fig.
1(b). This function gives the S-form of the static \textit{I}(\textit{V})
characteristic of the plasma diode.

Under these circumstances, from the circuit equations [$U=i_{1}r+u$; $%
i_{1}=i+C\dfrac{du}{dt}$;\linebreak\ $u=L\dfrac{di}{dt}+f(i)$], by
eliminating $i_{1}$, we get%
\begin{equation}
\left\{ 
\begin{array}{l}
\dfrac{di}{dt}=-\dfrac{1}{L}f(i)+\dfrac{1}{L}u\equiv F(i,u), \\ 
\dfrac{du}{dt}=\dfrac{U}{rC}-\dfrac{1}{C}i-\dfrac{1}{rC}u\equiv G(i,u).%
\end{array}%
\right.   \label{eq.1}
\end{equation}%
The \textit{rhs} in eqs. (\ref{eq.1}) give just the reaction functions 
\textit{F} and \textit{G} for the plasma system analyzed here. Adding the
appropriate diffusion terms, the RD (or drift-diffusion) system of equations
reads:%
\begin{equation}
\left\{ 
\begin{array}{l}
\dfrac{\partial i}{\partial t}=F(i,u)+D_{i}\nabla ^{2}i, \\ 
\dfrac{\partial u}{\partial t}=G(i,u)+D_{u}\nabla ^{2}u.%
\end{array}%
\right.   \label{eq.2}
\end{equation}%
Here \textit{D}$_{i}$ accounts for the diffusion of the positive ions and 
\textit{D}$_{u}$ for that of electrons \cite{Sandu1}.

In the stationary homogeneous state the solutions of the system are the
fixed points [$F(i,u)=0$ and $G(i,u)=0$] given by%
\begin{equation}
f(i_{0})=u_{0}=U-ri_{0}.  \label{eq.3}
\end{equation}

Since the stationary homogeneous state of the plasma system (\textit{i.e.}
when the ball of fire is not formed yet on the anode surface, but the system
is on the lower branch of the bistability region) is stable, this agrees
with the first condition required for the emergence of Turing structures:
the stationary homogeneous state is stable to small homogeneous
perturbations (\textit{i.e.} in the absence of diffusion). By imposing this,
we can find the conditions that must be satisfied by the system's equations
when small perturbations act upon it (Lyapunov method). Let the perturbed
values of $i_{0}$ and $u_{0}$ be $i=i_{0}+x$, $x<<i_{0}$ and $u=u_{0}+y$, $%
y<<u_{0}$. As long as the fixed points of the system of equations are not
hyperbolic (\textit{i.e.} they are not bifurcation points), Hartman-Grobman
theorem from Nonlinear Dynamics \cite{Nayfeh} is valid: "In an open
neighborhood around a nonhyperbolic fixed point, the dynamics of the
nonlinear system is topologically equivalent with the dynamics of the
linearized system". So, linearizing the static $I(V)$ characteristic of the
system around the fixed point $(i_{0},u_{0})$, we get:\linebreak\ $%
f(i)=f(i_{0})+\left. \frac{df}{di}\right\vert _{i=i_{0}}\cdot \left(
i-i_{0}\right) +...\cong f(i_{0})+R_{d}\cdot x$, where $\left. \frac{df}{di}%
\right\vert _{i=i_{0}}=R_{d}$ is the dynamic or differential resistance of
the nonlinear circuit element in the working point $(i_{0},u_{0})$.

In this way eq. (\ref{eq.1}) can be written%
\begin{equation}
\left\{ 
\begin{array}{c}
\dfrac{dx}{dt}=-\dfrac{R_{d}}{L}x+\dfrac{1}{L}y, \\ 
\dfrac{dy}{dt}=-\dfrac{1}{C}x-\dfrac{1}{rC}y.%
\end{array}%
\right.   \label{eq.4}
\end{equation}

One necessary condition for the emergence of Turing structures is the
presence of cross-inhibition \cite{Szili}. Analyzing eq. (\ref{eq.4}), the
cross-inhibition is present if the products of the diagonal elements of the
Jacobian matrix of the above system are negative \cite{Szili}:%
\begin{equation}
\left\{ 
\begin{array}{c}
\left( -\dfrac{R_{d}}{L}\right) \cdot \left( -\dfrac{1}{rC}\right) <0, \\ 
\left( \dfrac{1}{L}\right) \cdot \left( -\dfrac{1}{C}\right) <0.%
\end{array}%
\right.   \label{eq.5}
\end{equation}%
If the second inequality is evident, the first one is satisfied if and only
if the differential resistance of the gaseous conductor is negative%
\begin{equation}
R_{d}<0.  \label{eq.6}
\end{equation}%
Indeed, this result confirms the experimental finding according to which the
ball of fire acts as a negative differential resistance in the diode's
circuit. It must be stressed here that, although the above analysis is made
for the stationary homogeneous state (\textit{i.e.} when the ball of fire is
not yet formed at the anode surface), the system is in the bistability
region, where the third, unstable state also exists and it displays a
negative differential resistance. Also, if the system is on the upper branch
of the bistability region, for the same value of the potential drop on the
plasma diode (\textit{i.e.} the control parameter of the system) as above,
the ball of fire will exist at the surface of the anode.

Taking into account the above analysis, the matrix of signs associated with
the Jacobian matrix of the system, evaluated in the stationary homogeneous
state, reads%
\[
sgnJ_{0}=\left( 
\begin{array}{cc}
+ & + \\ 
- & -%
\end{array}%
\right) . 
\]%
This corresponds to a model of Turing structure called activator - substrate
depleted \cite{Engel} and is of the same type as the Brusselator \cite{Koch}%
. In the frame of this model the substrate is depleted during the
autocatalysis process. Its consumption slows down the auto-amplification
process of the activator concentration. Activator - substrate depleted
systems are characterized by opposite signs for the couplings of a species
with itself and with the other one. The result is a spatial distribution of
concentrations in opposition of phase: in the place where one species
concentration is maximal, the others concentration is minimal and vice
versa. The concentration peaks for this type of system are not sharp, but
rounded, the new peaks being formed by the splitting of the existing ones
and their shifting \cite{Koch}. The explanation is as follows: the substrate
concentration is larger in the vicinity of the place where the activator
peak grows by depleting the substrate. This determines the activator
concentration to grow in lateral direction. As a result, the concentration
peak of the activator will split up and shift towards the regions that have
a higher concentration of substrate. This scenario explains the formation of
several balls of fire on the anode \cite{Ionita,Aflori} as well as
their mutual disposition on the anode, in the form of regulate polygons \cite%
{Ivan}.

The stability of eqs. (\ref{eq.2}) to small homogeneous perturbations is
ensured if the trace and the determinant of the Jacobian matrix $J_{0}$ of
eqs. (\ref{eq.4}), evaluated in the stationary state, satisfies
concomitantly the following conditions \cite{Murray}:%
\begin{equation}
\left\{ 
\begin{array}{c}
TrJ_{0}<0, \\ 
DetJ_{0}>0,%
\end{array}%
\right.   \label{eq.7}
\end{equation}%
or%
\begin{equation}
\left\{ 
\begin{array}{c}
\left\vert R_{d}\right\vert <\dfrac{L}{rC}=\dfrac{Z_{0}^{2}}{r}, \\ 
\left\vert R_{d}\right\vert <r,%
\end{array}%
\right. ,  \label{eq.8}
\end{equation}%
where $Z_{0}^{2}=L/C$ is the proper impedance of the ball of fire. In
conclusion, one necessary condition for the appearance of Turing structures
is:%
\begin{equation}
\left\vert R_{d}\right\vert <\min \left\{ r;\dfrac{Z_{0}^{2}}{r}\right\} .
\label{eq.9}
\end{equation}

Since $Z_{0}$ always lies between $r$ and $Z_{0}^{2}/r$, this means that the
stationary ball of fire permanently self-adjusts its structure such that its
negative differential resistance to be smaller than its proper impedance.
The fact that the negative differential resistance of the ball of fire is
smaller than $Z_{0}^{2}/r$ means that any small perturbation acting on the
system extinguishes, the system behaving as a damped oscillator. The
equation for this oscillator can be easily obtained by eliminating $u$ from
the second eq. (\ref{eq.1}) with the help of the last circuit eq. and the
damping coefficient of this oscillator is $\delta =\frac{1}{2L}\left( \frac{%
Z_{0}^{2}}{r}-\left\vert R_{d}\right\vert \right) $. The perturbations'
damping is ensured as long as the potential drop on the plasma diode does
not surpass another critical value, at which the ball of fire disrupts \cite%
{Sandu1}. Above this new critical value, the current variations become
periodic (\textit{i.e.} $\left\vert R_{d}\right\vert =Z_{0}^{2}/r$ or $%
\delta =0$) and the double layer bordering the ball of fire periodically
detaches from its surface and travels a certain distance towards the
cathode. This new critical value of the potential drop on the plasma diode
delimitates the border between spatial and spatiotemporal self-organization.

The second necessary condition for the appearance of Turing structures
requires that the stationary homogeneous state to be unstable to
inhomogeneous perturbations (\textit{i.e.} in the presence of diffusion).
With the notation $\alpha =D_{u}/D_{i}$, the mathematical conditions
expressed by the above statement are \cite{Murray}:%
\begin{equation}
\left\{ 
\begin{array}{ccc}
\alpha F_{x}+G_{y} & > & 0, \\ 
\left( \alpha F_{x}+G_{y}\right) ^{2}-4\alpha \det J_{0} & > & 0,%
\end{array}%
\right.   \label{eq.10}
\end{equation}%
where $F_{x}\equiv \frac{\partial F}{\partial x}$ and $G_{y}\equiv \frac{%
\partial G}{\partial y}$, respectively. Eqs. (\ref{eq.7}) and (\ref{eq.10})
define the so-called Turing domain, \textit{i.e.} the domain in the
parameters' space in which the appearance of Turing structures is possible.
Solving eqs. (\ref{eq.10}) we get:%
\begin{equation}
\left\{ 
\begin{array}{c}
\left\vert R_{d}\right\vert >\dfrac{Z_{0}^{2}}{\alpha r}, \\ 
\left\vert R_{d}\right\vert >\dfrac{Z_{0}^{2}}{\alpha r}\left( \dfrac{2r%
\sqrt{\alpha }}{Z_{0}}-1\right).%
\end{array}%
\right.   \label{eq.11}
\end{equation}

From the first eq. (\ref{eq.8}) and the first eq. (\ref{eq.11}) a very well
known result from the theory of Turing structures can be obtained: $\alpha >1
$. With other words, in any Turing system the activator is always diffusing
slower than the inhibitor or substrate depleted (\textit{i.e.} electrons
diffuse faster than positive ions).

Eqs. (\ref{eq.11}) can be rewritten as follows:%
\begin{equation}
\left\vert R_{d}\right\vert >\max \left\{ \dfrac{Z_{0}^{2}}{\alpha r};\dfrac{%
Z_{0}^{2}}{\alpha r}\left( \dfrac{2r\sqrt{\alpha }}{Z_{0}}-1\right) \right\}.
\label{eq.12}
\end{equation}

Reuniting eqs. (\ref{eq.9}) and (\ref{eq.12}), the variation interval for
the negative differential resistance of the ball of fire can be written as
follows:%
\begin{equation}
\max \left\{ \dfrac{Z_{0}^{2}}{\alpha r};\dfrac{Z_{0}^{2}}{\alpha r}\left( 
\dfrac{2r\sqrt{\alpha }}{Z_{0}}-1\right) \right\} <\left\vert
R_{d}\right\vert <\min \left\{ r;\dfrac{Z_{0}^{2}}{r}\right\}.  \label{eq.13}
\end{equation}

\section{Conclusions}

Establishing the basic equivalent electric circuit for a plasma diode, at
the anode of which a ball of fire is formed, the reaction functions of the
RD equations system can be easily derived. Applying the Turing formalism to
that system of equations, it is shown that the ball of fire belongs to the
same class of Turing structures like the Brusselator in chemistry. Moreover,
one of the necessary conditions for the existence of the ball of fire at the
anode surface in a plasma diode, resulting from experiments, is the negative
sign of its differential resistance. At this conclusion also arrives the
present approach, as a consequence of applying the Turing mathematical
formalism. The variation interval for the negative differential resistance
of the ball of fire is also derived by imposing all the conditions defining
the Turing domain in the parameters space.

\acknowledgments This work was supported by the Romanian Ministry of
Education and Research, Grant No. 33373/29.06.2004, topic 9, code CNCSIS 74.


\begin{thebibliography}{10}
\bibitem[1]{Turing} Turing A. M., \textit{Phil. Trans. Roy. Soc. }B \textbf{%
237} (1952) 37.

\bibitem[2]{Prigogine1} Prigogine I., Nicolis G., \textit{J. Chem. Phys. }%
\textbf{46} (1967) 3542.

\bibitem[3]{Prigogine2} Prigogine I., Lefever R., \textit{J. Chem. Phys. }%
\textbf{48} (1968) 1695.

\bibitem[4]{Kepper} Casets V., Dulos E., Boissonade J., de Kepper P., 
\textit{Phys. Rev. Lett. }\textbf{64} (1990) 2953.

\bibitem[5]{Ouyang} Ouyang Qi, Swinney H. L., \textit{Nature} \textbf{352}
(1991) 610.

\bibitem[6]{Murray} Murray J. D., \textit{Mathematical Biology}
(Springer-Verlag, Berlin-Heidelberg-New York) 1989.

\bibitem[7]{Koch} Koch A. J., Meinhardt H., \textit{Rev. Mod. Phys. }\textbf{%
66} (1994) 1481.

\bibitem[8]{Kerner} Kerner B. S., Osipov V. V., \textit{Autosolitons: A New
Approach to Problems of Self-Organization and Turbulence} (Kluwer Academic
Publishers, Berlin) 1994.

\bibitem[9]{Lepp} Lepp\"{a}nen T. \textit{et al.}, \textit{Brazilian J. Phys.%
} \textbf{34} (2004) 368.

\bibitem[10]{Astrov} Astrov Y. \textit{et al.}, \textit{Phys. Lett. }A 
\textbf{211} (1996) 184.

\bibitem[11]{Purwins} Astrov Y., Ammelt E., Purwins H. G., \textit{Phys.
Rev. Lett.} \textbf{78} (1997) 3129.

\bibitem[12]{Loz1} Lozneanu E., Popescu S., Sanduloviciu M., IEEE \textit{%
Trans. Plasma Sci.} \textbf{30} (2002) 32.

\bibitem[13]{Loz2} Lozneanu E., Popescu V., Popescu S., Sanduloviciu M.,
IEEE \textit{Trans. Plasma Sci.} \textbf{30} (2002) 30.

\bibitem[14]{Popescu} Popescu S., \textit{Analele Universit\u{a}\c{t}ii
\textquotedblleft Al.I.Cuza\textquotedblright\ Ia\c{s}i} \textbf{XLIX}
(2003) 95.

\bibitem[15]{Vastano} Vastano J. A. \textit{et al.}, \textit{J. Chem. Phys. }%
\textbf{88} (1988) 6175.

\bibitem[16]{Szili} Szili L., Toth J., \textit{Phys. Rev. }E \textbf{48}
(1993) 183.

\bibitem[17]{Sandu1} Sanduloviciu M., Lozneanu E., Popescu S., \textit{%
Chaos, Solitons and Fractals} \textbf{17} (2003) 183.

\bibitem[18]{Leu} Sanduloviciu M., Borcia C., Leu G., \textit{Phys. Lett. }A 
\textbf{208} (1995) 136.

\bibitem[19]{Popescu2} Popescu S., Lozneanu E., \textit{J. Plasma and Fusion
Res.} SERIES- Japan \textbf{4} (2001) 559.

\bibitem[20]{Nayfeh} Nayfeh A. H., Balachandran B., \textit{Applied nonlinear
dynamics -- analytical, computational and experimental methods} (John Willey
\& Sons, Inc., New York) 1995.

\bibitem[21]{Engel} Engelhardt R., \textit{Modelling Pattern Formation in
Reaction -- Diffusion Systems}, Disertation Thesis (H. C. Oersted Institute,
Univ. of Copenhaga) 1995.

\bibitem[22]{Ionita} Ionita C., Dimitriu D. G., Schrittwieser R., \textit{%
Int. J. Mass Spectrom.} \textbf{233} (2004) 343.

\bibitem[23]{Aflori} Aflori M. \textit{et al.}, IEEE \textit{Trans. Plasma
Sci.} \textbf{33} (2005) 542.

\bibitem[24]{Ivan} Ivan L. M. \textit{et al.}, IEEE \textit{Trans. Plasma
Sci.} \textbf{33} (2005) 544.
\end{thebibliography}
\end{document}